# Do ROS really slow down aging in *C. elegans*?

Yaguang Ren[a,b,1,*], Sixi Chen[b,1], Mengmeng Ma[a,b], Congjie Zhang[b], Kejie Wang[b], Feng Li[b], Wenxuan Guo[a], Jiatao Huang[b], Chao Zhang[a,*]

[a] Translational Medical Center for Stem Cell Therapy and Institute for Regenerative Medicine, Shanghai East Hospital, Shanghai Key Laboratory of Signaling and Disease Research, School of Life Sciences and Technology, Tongji University, Shanghai, 200092, China
[b] Zhejiang Provincial Key Laboratory of Medical Genetics, School of Laboratory Medicine and Life Sciences, Wenzhou Medical University, Wenzhou, Zhejiang, 325035, China
[1] These authors contributed equally to this work.
*Correspondence: renyaguang3@163.com (Y. R); zhangchao@tongji.edu.cn (C. Z)

**Abstract**

  It was shown in *C. elegans* that life-long treatments with low level of the prooxidant paraquat increased reactive oxygen species (ROS), which then acted as the signal to slow down aging by activating the intrinsic apoptosis pathway. And deficiencies of the electron transport chain subunits increased longevity through similar mechanisms. We surprisingly found that ROS were down- and up-regulated respectively upon chronic and transient paraquat treatments although the same ROS measuring approaches were applied. Adaptive responses were shown to be responsible for the reduction under chronic stresses. Post young adult stage started treatments shortened lifespan at all concentrations tested and mild extension was only observed by treating with "proper" concentration of paraquat from egg hatching. However, the extension in the latter case could be explained by growth retardation. Thus, sometimes the extension of the post growth span rather than the whole lifespan should be considered as evidence for "real" slowing down of aging. We also believe that aging was affected by the synthetic effects of multiple secondary responses rather than that of ROS alone. If the protective elements such as antioxidants are beneficial then ROS should be not considering that the former fight against the latter. The view that ROS are "good" for aging would also produce a false impression that antioxidants are bad. We hope the alternative perspective proposed here will bring about beneficial discussions in correlated areas.

**Highlights**

- Transient and chronic prooxidant treatments up- and down-regulate ROS respectively.
- Adaptive responses were responsible for ROS reduction under chronic stresses.
- Protective mechanisms activated rather than ROS are beneficial for longevity.
- Post growth span rather than the whole lifespan should be used sometimes.

## Graphical Abstract

The synthetic effects of secondary responses rather than that of ROS alone affect aging

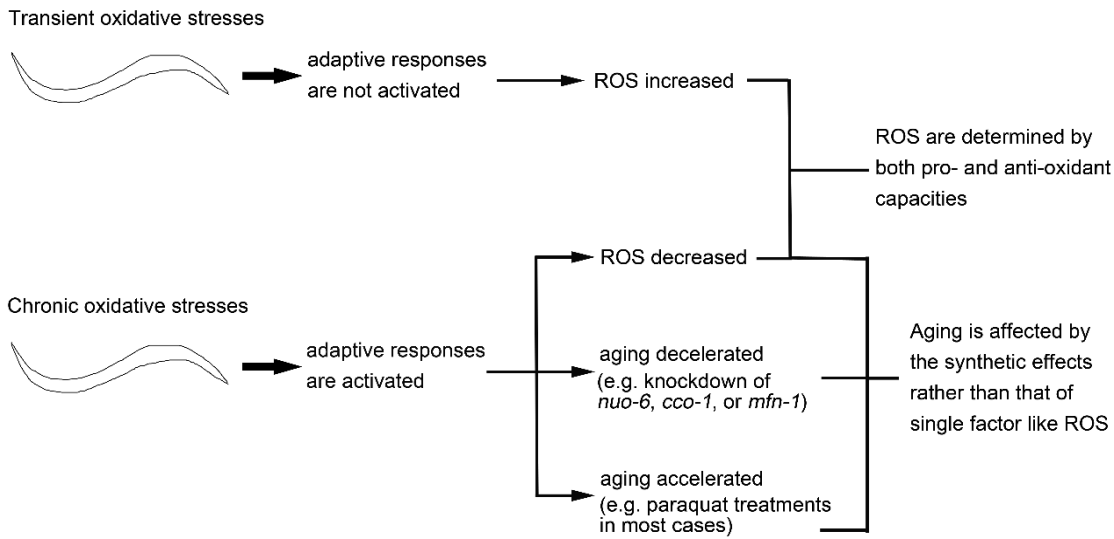

## 1. Introduction

The oxidative stress theory of aging states that cellular damages caused by reactive oxygen species can lead to aging [1]. As organism ages ROS increase and oxidative damages such as protein carbonization and lipid peroxidation also accumulate [2, 3]. The discovery of SODs, catalases, glutathione S-transferase and other antioxidants suggests that organisms have developed elegant mechanisms during evolution to tackle against the persistently generated ROS [4]. Many studies showed that the antioxidant resveratrol (RSV) was able to slow down aging in both invertebrates and vertebrates [5, 6]. Genetic or environmental perturbations that postpone aging such as reduced insulin/IGF-1 signaling (IIS), mitochondrial dysfunctions, and dietary restriction (DR) usually activate the expression of antioxidant enzymes and enhance oxidative stress resistance in *C. elegans* [7, 8]. The DAF-16/FoxO3a-dependent longevity signal was shown to be initiated by antioxidants [9]. Knockout of the worm thioredoxin *trx-1* shortened and overexpression of it mildly prolonged lifespan [10]. These studies are to some extent in accordance with the oxidative stress theory of aging.

However, recently the view that ROS are beneficial and can slow down aging is getting popular [11]. This counterintuitive conclusion is largely based on reports that chronic treatments with proper concentration of the prooxidant paraquat (PQ) in *C. elegans* slowed down aging by increasing ROS [12, 13], and deletion of all five SODs in *C. elegans* did not decrease longevity [14]. Moreover, the intrinsic apoptosis signaling pathway (CED-9/Bcl-2, CED-4/Apaf1, and CED-3/Casp9) activated by mitochondrial ROS was considered as the reason for lifespan extension [13]. However, we have reasons to question that conclusion. Firstly, the retardation of growth was overlooked when translating lifespan date into aging. For example, if growth retardation was taken into account deletion of all five SODs in *C.elegans* would shorten lifespan. Secondly, although good reliability of ROS detecting approaches is prerequisite for getting any scientific conclusions, fluorescent images of worms stained with ROS probing dyes were not shown [12, 13]. Thirdly, ROS are usually measured in isolated mitochondria which are reported to be in unphysiological state [15]. Fourthly, although apoptosis signaling pathway was considered as the reason for slowing down of aging no signs of apoptosis was found [13], which was somewhat self-contradictive. Fifthly, ROS in isolated mitochondria of *isp-1(qm150)* and *nuo-6(qm200)* mutants were previously shown to be decreased [12]. However, the same research group later reported increased ROS in the same strains [13]. It was again self-contradictive and no explanation was given. Sixthly, lost or abnormally died worms were replaced with those from backup plates during lifespan tests in previous studies [13], which was different from the standard method [16, 17]. And the manual selection of worms may lead to artificial results. Finally, multiple protective responses including antioxidant enzymes were activated under prooxidant stresses [8]. Although there were other studies emphasizing the beneficial role of ROS in aging [18, 19], we believe that the pro-longevity effects, if any, should be ascribed to the synthetic effects of secondary responses rather than that of ROS.

In this study, we found that transient and chronic PQ treatments resulted in increased and decreased ROS respectively and the activation of protective mechanisms was responsible for ROS reduction in the latter case. This result is in consistent with previous report that chronic treatment with 0.25 mM PQ led to decreased ROS in *C. elegans* [19]. We further found that the longevity was shortened by young adult stage started treatments at all concentrations tested. Although mild increase of longevity was found at proper concentration upon treating from egg hatching, this should be explained by synthetic effects of secondary responses such as retardation of growth, activation of antioxidant enzymes, and other unknown factors. Furthermore, deficiency of *nuo-6* or *cco-1* resulted in decreased rather than increased ROS levels. In combination with previous studies we here question the view that ROS are beneficial for aging.

## 2. Materials and methods

2.1. *C. elegans* strains and maintenance

The worm strains were obtained from Caenorhabditis Genetics Center (University of Minnesota, USA), and were grown on E. coli OP50 seeded NGM plates at 20°C, as described by Brenner [20]. The following strains were used: N2 (wide type), SJ4100 *zcIs13 [hsp-6::GFP]*, CF1553 *muIs84 [(pAD76) sod-3p::GFP + rol-6],* GA480 *sod-2(gk257) I*, *sod-3(tm760) X*, CF1038 *daf-16(mu86) I*, MQ1333 *nuo-6(qm200) I*.

2.2. Paraquat or juglone treatments

NGM plates containing 0, 0.03, 0.05, 0.1, 0.2, 0.3, and 0.5 mM PQ (Sigma Aldrich) were prepared by adding 0, 90, 150, 300, 600, 900, and 1500 μl of stock solution (0.1 M) into 300 ml of liquid NGM respectively, and were mixed well just before being poured into ten 6-cm-diameter plates. The plates were stored at 4°C and were used within two weeks. For egg hatching started treatments, about twenty 1-day old adults were transferred onto the OP50 lawn of PQ plate respectively (plates without PQ were included as control), and were removed after eight to ten hours of egg laying. Three days later, 60 synchronized progenies in each group were transferred onto fresh prepared plates containing corresponding concentrations of PQ, then were transferred each day during egg laying period and every other day after that until being tested in order to avoid starvation. Young adult stage started treatments were carried out similarly. For transient treatments 50 synchronized 1-day old adults were placed onto the OP50 lawn of the PQ plates and were maintained at 20°C for 30 minutes. Treatments using the prooxidant juglone (Sigma aldrich) were performed through similar procedures, except that the final concentrations were 10, 20, 40, 80, and 160 μM respectively.

2.3. ROS and superoxide measurements

ROS measurement in living worms was carried out as described with minor modifications [21, 22]. At each time point 20 synchronized worms from each group were transferred into 200 μl of M9 buffer containing 10 μM H2DCFDA in an Eppendorf tube and were incubated in the dark for 1.5 hours. Then the worms were placed in 10 μl of M9 buffer containing 0.4 mM levamisole for anesthesia on the thin dried agarose pad above the glass slide, and were mounted onto the Nikon Ni-U fluorescence microscope for capturing of the green fluorescence of the oxidized product dichlorofluorescin (DCF). Autofluorescence of simultaneously cultured unstained worms was also captured. For better visualization the contrast and intensity of the micrographs from control and tested groups were modified simultaneously using Photoshop CS4 software after being merged into one large image. For quantitatively comparing relative ROS and superoxide levels original micrographs (those unmodified) were used. Background fluorescence and autofluorescence intensities were subtracted when comparing relative ROS levels among samples. For measuring ROS in body lysates worms were collected and washed three times in S buffer to get rid of bacteria, and then were homogenized on ice. After centrifugation at 12000 g for twenty minutes, 100 μl of the supernatant was mixed with fresh prepared 100 μl of 20 μM H2DCFDA in S buffer, and were transferred into 96-well plate. The fluorescence intensity was measured using the fluorescence microplate reader every 20 min for up to 2 h with an excitation 488 nm and an emission 511 nm. Background fluorescence intensity measured in S buffer was subtracted. ROS levels were normalized to protein concentration of the supernant. ROS measurement in isolated mitochondria were performed according to previous methods [12, 23]. Superoxide levels in living worms, body lysates, and isolated mitochondria were analyzed similarly, except that the dyes mitosox (Life technologies) (excitation at 510 nm and emission at 580 nm) or DHE (Beyotime) (excitation at 300 nm and emission at 610 nm) were used at the final concentration of 5 μM.

## 2.4. Lifespan tests

Around 10 gravid adults were put onto the corresponding plates seeded with OP50 strain and were let lay eggs for 6 to 8 hours before being removed. After reached young adult stage, 80 to 100 F1 generation worms were transferred onto fresh prepared plates every day during the reproduction period, and every two or three days after that. Worms that did not respond to gentle nose touch with the toothpick were considered dead, and that crawled off the plate or died from hatched embryos in the uterus were scored as censored data.

## 2.5. Phenotype analysis

Phenotypes including body size, pharyngeal pumping rate, and brood size were analyzed as described [24, 25].

## 2.6. Stress resistance assay

The stress resistance assays were carried out according to previous methods [24, 26, 27].

## 2.7. Ultraviolet irradiation

Around thirty synchronized young adults were transferred onto fresh prepared NGM plate and were exposed to low dose of ultraviolet radiation ($100J/m^2$) in a 254 nm-UV crosslinker (UVP). A simultaneously prepared group of worms not irradiated were used as control.

## 2.8. Short term heat shock

Around thirty synchronized young adults were transferred onto OP50 seeded NGM plate, and were maintained at 37°C for 30 minutes. A simultaneously prepared group of worms not heat-shocked were used as control.

## 2.9. Real-time PCR

Total RNAs were isolated using Trizol reagent (Invitrogen) according to manufacturer's instructions. The mRNA was reverse transcribed to cDNA and quantitative Real-time PCR was performed as previously described [28]. The level of *act-1* mRNA was used as internal control. Primer sequences were shown in Table S2.

## 2.10. RNAi feeding experiments

The E. coli strain HT115 (DE3) containing the corresponding RNAi construct was inoculated in LB medium containing 50 µg/ml ampicillin and let grow overnight at 37°C and 250 rpm, then IPTG was added to induce the production of dsRNA (the final concentration of IPTG was 1 mM). After that 100 µl of the medium was poured onto the NGM plates containing 50 µg/ml ampicillin and 1 mM IPTG, which were then maintained at 25°C for 12 to 24 hours. 10 gravid adults were put onto the bacterial lawn of the corresponding plates and were let lay eggs for 10 hours. The adults were then removed and the synchronized F1 generation worms were used for further analysis.

2.11. Quantification and statistical analysis

The Kaplan-Meier method and log-rank test were performed for analysis of survival curves in lifespan measurements and stress resistance tests. The Kruskal–Wallis (non-parametric) test was used for analyzing the effects of PQ treatments on growth. The Mann Whitney test was used for comparing progeny numbers between each treated group and control. Student's-t test was used for comparing data from Real-time PCR experiments and other analyses. Differences in means were considered statistically significant at $p < 0.05$. Significance levels are: * $p < 0.05$; ** $p < 0.01$; *** $p < 0.001$. The Graphpad Prism 5.0 software was used for statistical analysis.

## 3. Results

3.1. ROS and superoxide levels are reduced by chronic PQ treatments

The dye 2′, 7′-dichlorofluorescin diacetate (H2DCFDA) is commonly used as the fluorescence-based probe to detect general ROS in vitro and in vivo [29-31], and has also been used for this purpose in *C. elegans* [12, 21]. The dyes dihydroethidium (DHE) and mitosox are usually used to analyze superoxide levels [12, 32]. In this study, these dyes were used for detecting ROS and superoxide in living worms, body lysates, and isolated mitochondria.

To explore the effects of chronic PQ treatments on ROS and superoxide generation we grew worms from egg hatching on OP50 seeded NGM plates containing 0, 0.03, 0.05, 0.1, 0.2, 0.3, and 0.5 mM PQ respectively, and found increased ROS and superoxide levels during aging in all groups. However, at the same chronological age worms treated with 0.1 mM or higher concentrations of PQ had much lower ROS and superoxide levels, compared to control (Fig. 1A–D). ROS and superoxide in gently prepared body lysates and isolated mitochondria were also found to be down-regulated at 0.2 mM (Fig. 1 E and F). Juglone, another well-known prooxidant [33, 34], led to similar ROS alternations (Fig. S1). These results are a strong blow to the view that PQ slows down aging through elevating ROS levels. Consistently, it was reported that increased ROS levels due to increased respiration activated antioxidant enzymes and led to further decrease of ROS in the long term [35]. Furthermore, adaptation to hydrogen peroxide enhances PC12 cell tolerance against oxidative damage [36].

**Fig. 1. Chronic PQ treatments resulted in reduced ROS and superoxide levels in living worms, body lysates, and isolated mitochondria.** (A, B) ROS and superoxide were revealed in living worms by staining with H2DCFDA (10 μM) and DHE (5 μM) respectively. Scale bars, 300 μm. (C, D) ROS and superoxide levels were normalized to that of 60-hour and 120-hour (t=0 at egg hatching) old worms from the control group respectively. Data are presented as mean ± SEM, and n

represents number of worms. (E, F) Treatments with PQ from egg hatching resulted in reduced ROS and superoxide levels in body lysates and isolated mitochondria respectively. ROS and superoxide were measured in worms at young adult stage (around 3.5-day old from egg hatching). Data are presented as mean ± SEM of at least three independent experiments.

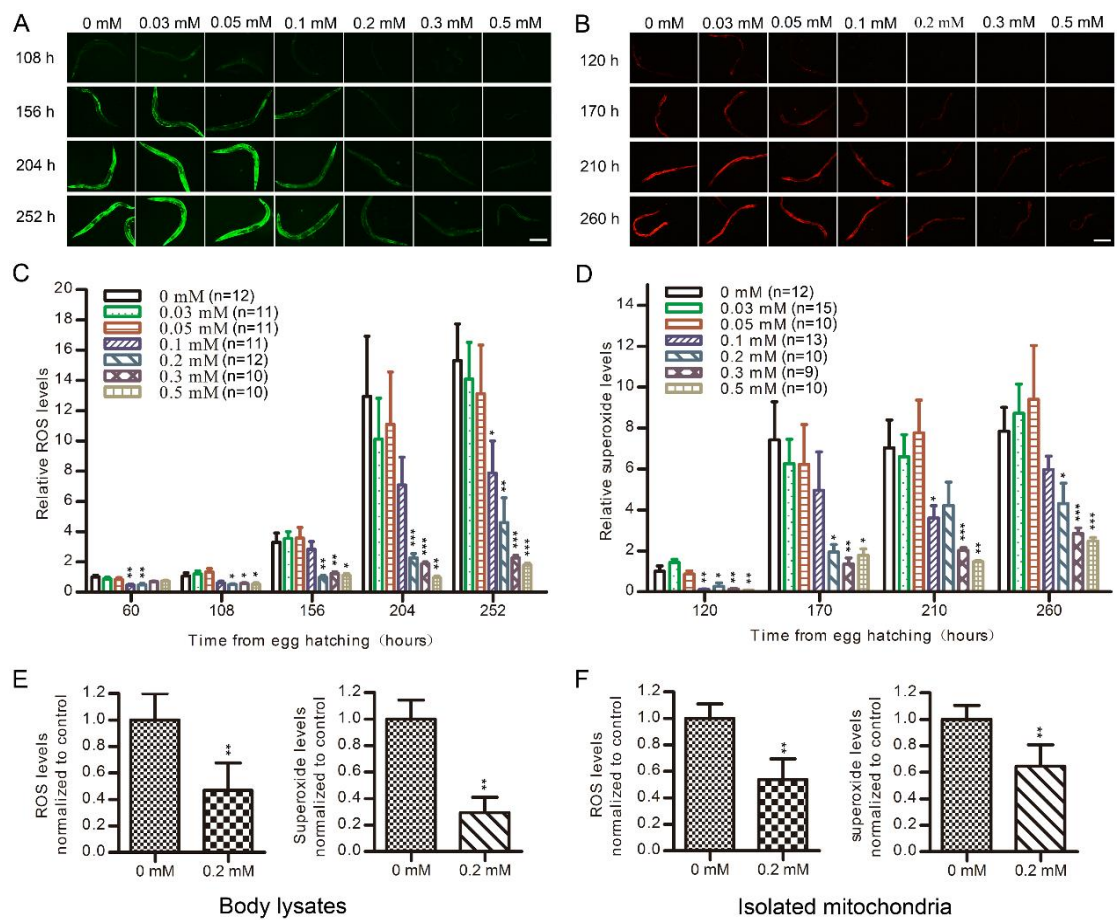

3.2. Chronic PQ treatments could not prolong lifespan in most cases, especially when growth retardation was taken into account

Worms were treated from egg hatching with gradually increased levels of PQ respectively and the average lifespan was not significantly affected at 0.03, 0.05, 0.1, 0.2, and 0.5 mM, but at 0.3 mM it was increased by two days compared to that of control. The extension was mild but statistically significant (Fig. 2A and Table S1). Worms' growth was found retarded at 0.1 mM and higher concentrations, and at 0.5 mM it was so severely affected that the final body size was much smaller than normal (Fig. 2B). This was why 0.5 mM was used as the upper limit. Conversely, 0.03 and 0.05 mM were used as the lower limits in many tests as growth was not retarded at these levels. Interestingly, the final body size was slightly larger than that of control except at 0.5 mM, which was reproducible but the reason was unknown. The delay of growth fully explains lifespan extension at least in this study. For example, it took about 10.5 days for worms treated with 0.3 mM PQ to reach the maximum of body length, while for those from control group only 6.5 days were required. Therefore, although at 0.3 mM the average longevity was increased by around two days the survival time measured from growth completion was in fact shortened. The reproduction was also postponed in PQ stressed worms and the progenies produced were fewer, suggesting reduced fertility (Fig. 2C).

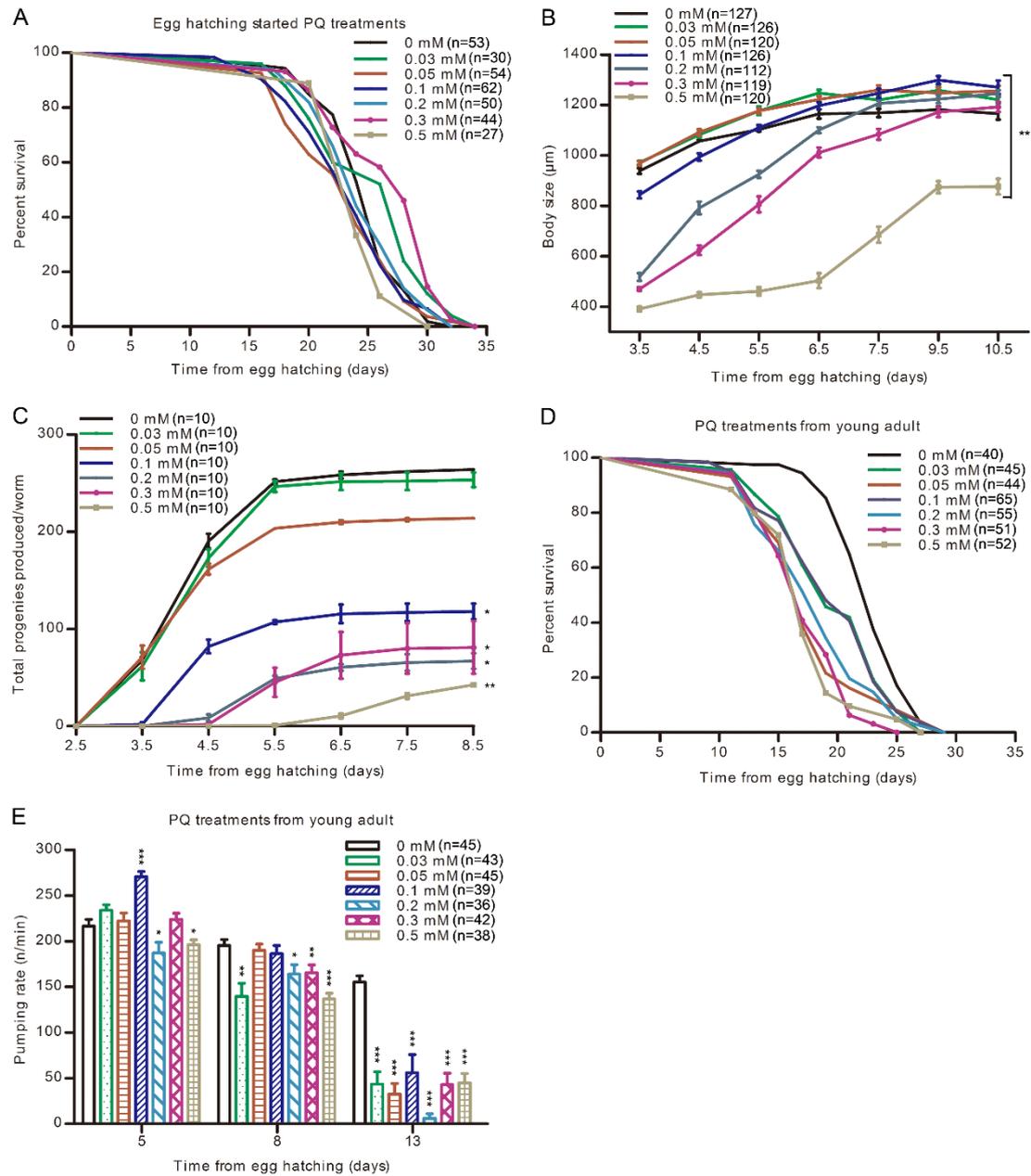

**Fig. 2. Chronic PQ treatments retarded physiological rate and could not prolong lifespan in most cases.** (A) Lifespan was mildly prolonged by treating with 0.3 mM PQ (p<0.05), but was not at other concentrations (p>0.05). Data shown are representative of three independent experiments. (B) Growth was retarded by chronically treating with 0.1mM or higher concentrations of PQ. (C) Chronic PQ treatments resulted in reduced and retarded reproduction. (D) Treating with PQ from young adult stage shortened lifespan. For each treatment compared to control, p<0.01. (E) PQ treatments from young adult stage accelerated aging associated pumping rate decline. n represents number of worms. See also Table S1.

Thus, retardation of physiological processes contributed largely if not completely to lifespan extension. It should be noted that the "proper" concentration shown here was 0.3 mM, similar to what was used (0.25 mM) previously [19]. However, in the study of Yee et al it was 0.1 mM and they also reported larger lifespan extension. These discrepancies may be explained by different experimental conditions, different methods used for lifespan analysis, or other unknown reasons. For example, we seeded OP50 bacteria onto paraquat containing NGM plates, while Yee et al transferred OP50 grown on regular NGM plates onto NGM-PQ plates using a platinum pick instead of seeding directly. They also replaced lost or abnormally died worms with those from a backup pool during lifespan tests, which was different from standard method [16, 17]. The manual selection is prone to produce artificial data, because even in the same cohort the individual worms' lifespan could be much different from each other and the lifespan of worms from backup plates could not match that of censored ones, especially during later stage of the test when worms became old and the large differences of locomotion activities among individuals making it difficult to decide which one to select for backup.

We speculate that if lifespan extension was largely due to growth retardation then the result would no longer be reproduced by treatments carried out from young adult stage. This was indeed the case, and what's more, shortened lifespan was observed at all concentrations tested (Fig. 2D and Table S1). Young adult started PQ treatments also accelerated aging associated pharyngeal pumping rate decline (Fig. 2E). Although we do not exclude the possibility that proper level of PQ could still increase longevity even if growth retardation is taken into account, just as shown in previous studies, the pro-longevity effect should not be ascribed to ROS but the synthetic effects of multiple secondary responses. Firstly, ROS are in fact decreased under chronic PQ treatments and the pro-longevity effect should not be ascribed to elevated ROS. Secondly, if the secondary responses are pro-longevity then ROS should be anti-longevity, because the persistently and excessively activated protective mechanisms such as the up-regulation of antioxidants function as countermeasures against ROS.

3.3. Adaptive mechanisms were activated upon chronic PQ treatments

Retrograde responses were shown to be activated under detrimental environments [8]. Consistently, we found that worms treated with PQ from egg hatching exhibited increased adaption to a variety of adverse conditions including hyperosmotic stress (Fig. 3A), extreme oxidative stress (Fig. 3B), and heat shock (Fig. 3C). For example, 1-day old adults from control group were paralyzed within eight minutes after being transferred onto NGM plates containing 500 mM NaCl, while some worms from 0.3 mM PQ treated group still kept moving 25 minutes later (Fig. 3A).

The enhanced adaption was not due to retardation of growth because the untreated larvae or young adults, whose body size are small, were still much sensitive to those stresses (our observation). The results suggested that adaptive responses should be activated under chronic PQ stresses. In support of this, 1-day old adults treated transiently with PQ for only thirty minutes exhibited increased ROS (Fig. 3D). The likely explanation should be that in such a short period the protective responses were not largely motivated because more time should be required for the transcription, translation, and maturation of antioxidant enzymes. Similar to transient PQ treatments, low dose of ultraviolet irradiation or short term of heat shock also led to ROS elevation (Fig. 3E), demonstrating the reliability of the ROS detecting method used in this study.

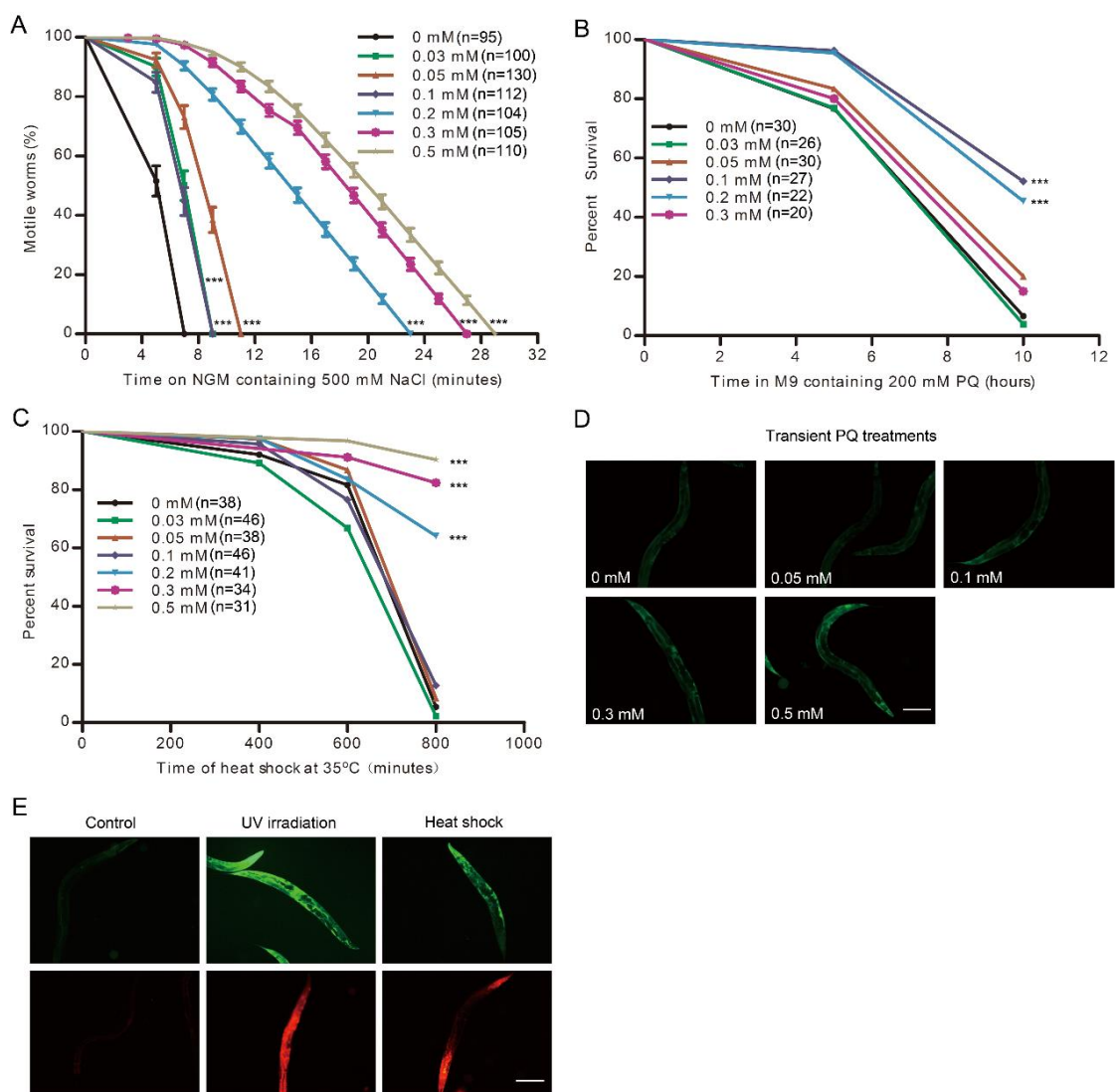

**Fig. 3. Adaptive responses were motivated under chronic PQ treatments.** (A-C) Chronic PQ treatments from egg hatching enhanced the resistance to high osmotic stress, extreme oxidative stress, and heat shock. n represents number of worms. (D) Transient treatments for only thirty minutes increased ROS levels. 1-day old adults (4.5-day old from egg hatching) were used for the assay. Scale bar, 200 μm. (E) As the positive control, low dose of UV irradiation or short term of heat shock increased ROS and superoxide levels. Synchronized young adults were used for the assay. Scale bar, 200 μm.

3.4. The elevated antioxidant capacity was responsible for ROS reduction under chronic PQ stresses

In response to oxidative stresses antioxidants and chaperones are activated [37, 38]. In this study, chronic PQ treatments also intensified the GFP fluorescence in CF1553 and SJ4100 strains (Fig. 4A), which carry the transgenic alleles *sod-3p::gfp* and *hsp-6p::gfp* respectively. Consistently, real-time PCR experiments revealed increased transcription of the superoxide dismutase *sod-3*, the glutathione S-transferase *gst-4*, and chaperones *hsp-6* and *hsp-16.2* (Fig. 4B).

We then investigated the role of these protective mechanisms in ROS regulation under chronic PQ treatments. N2 young adults were treated with PQ for two days and lower ROS levels were found compared to control (Fig. 4C). But in the *sod2/sod3* double mutant GA480 and the *daf-16* mutant CF1038 the reduction of ROS were not observed under similar treatments and in some cases ROS levels even increased (Fig. 4D and E). It was shown that the FOXO-family transcription factor *daf-16*, which acts downstream of the insulin/IGF-1 signaling pathway, played important roles in modulating oxidative stress response through modulating the expression of a variety of genes including antioxidant enzymes, chaperones, antimicrobial and metabolic genes [39]. The mitochondrial antioxidant *sod-3* was also reported to be regulated by *daf-16* [40]. Therefore, the results suggested that the superoxide dismutase *sod-2/3* and the transcription factor *daf-16* were essential for ROS regulation under chronic PQ stresses. Other protective genes such as *gst-4*, *hsp-6,* and *hsp-16.2* might also play some roles which deserve further investigations.

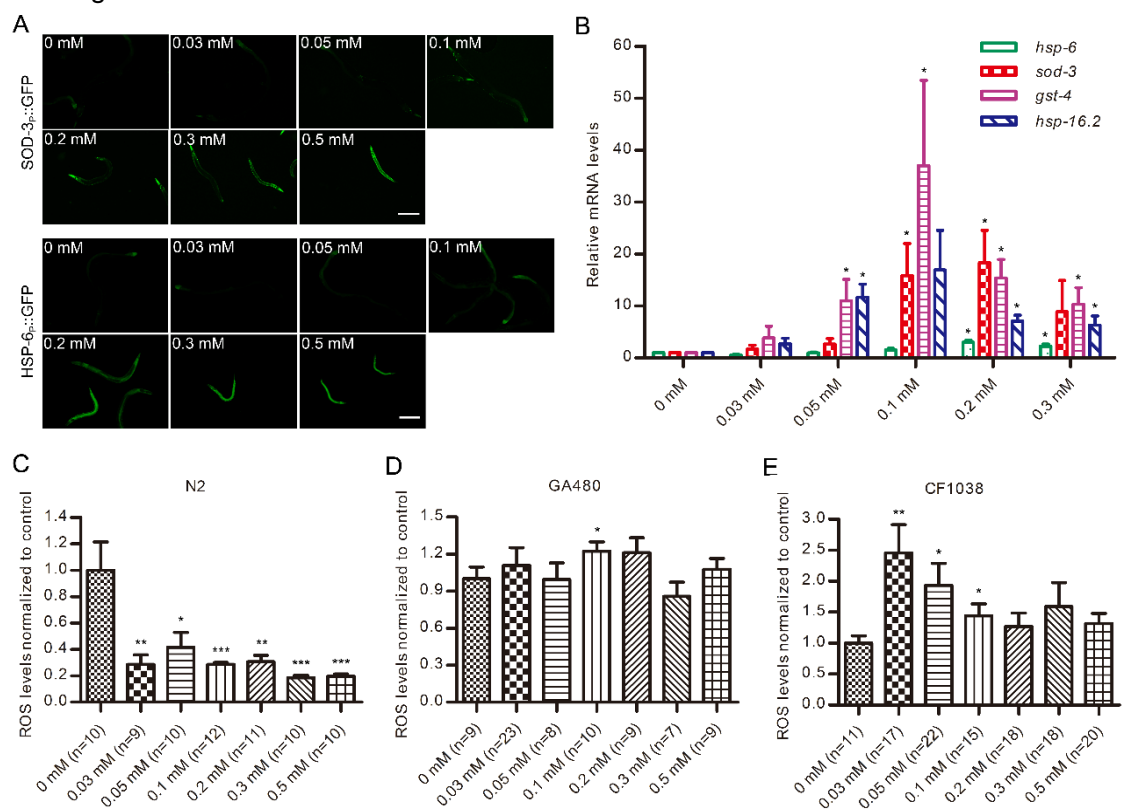

**Fig. 4. Protective mechanisms activated were responsible for ROS reduction upon chronic PQ treatments.** (A) *sod-3* and *hsp-6* were up-regulated by egg hatching started PQ treatments as shown by the enhanced fluorescence in reporter strains. Scale bars, 200 μm. (B) Real-time PCR experiments revealed increased transcription of *hsp-6*, *sod-3*, *gst-4,* and *hsp-16.2* genes in PQ stressed worms. (C) For N2 strain, after two days of young adult stage started PQ treatments ROS levels were lower, compared to that of control. ROS were measured in living worms. (D, E) For GA480

and CF1038 strains, ROS were no longer reduced under similar treatments. Data are presented as mean ± SEM. In C, D, and E, n represents number of worms.

3.5. Deficiencies of *nuo-6* or *cco-1* resulted in decreased ROS levels

Knockdown or knockout of the electron transport chain subunits *nuo-6*, *cco-1*, or *isp-1* were reported to increase longevity [41, 42], while *mev-1* or *gas-1* deficiencies decreased it [42, 43]. We found dramatically decreased ROS levels in *nuo-6* or *cco-1* RNAi knockdown worms (Fig. 5A). But ROS level in *mev-1* depleted worms was only slightly lower (Fig. 5A), compared to that of control. ROS and superoxide levels were further measured in gently isolated mitochondria of *nuo-6* RNAi worms and were also found to be decreased (Fig. 5B and C).

**Fig. 5. Knockdown of *nuo-6* or *cco-1* resulted in decreased ROS levels and up-regulated expression of SODs.**

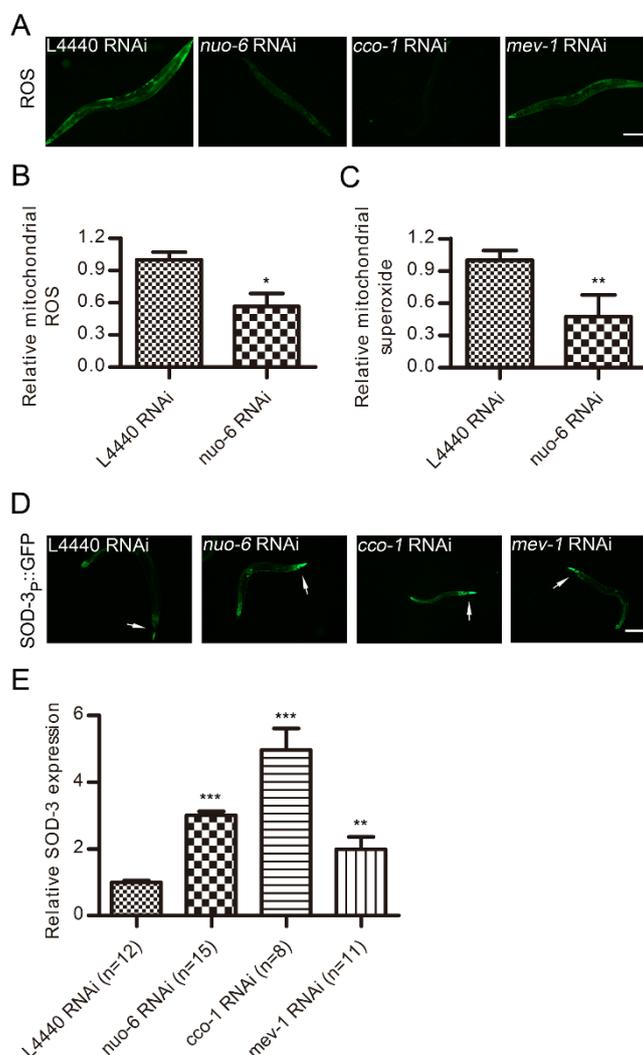

(A) ROS were dramatically down-regulated in *nuo-6* or *cco-1* RNAi treated 4-day old worms, compared to that in control (L4440 RNAi). But in *mev-1* RNAi worms ROS were not obviously affected. Scale bar, 200 μm. (B, C) *nuo-6* RNAi led to reduced ROS and superoxide levels in isolated mitochondria. 3-day old worms (t=0 at egg hatching) were used for the measurements. Data are presented as mean ± SEM of at least two independent experiments. (D) *nuo-6*, *cco-1*, or *mev-1* RNAi up-regulated the expression of *sod-3* as shown by the enhanced fluorescence in 4-day old reporter strain CF1553. Arrow indicates head region. Scale bar, 200 μm. (E) The fluorescence intensities in *nuo-6*, *cco-1*, or *mev-1* RNAi treated CF1553 worms were normalized to that in untreated control respectively. Data are presented as mean ± SEM, and n represents number of worms.

The expression of the superoxide dismutase *sod-3* was significantly up-regulated by RNAi treatment of *nuo-6*, *cco-1*, or *mev-1* as indicated by the enhanced fluorescence in the reporter strain CF1553, especially in the head regions (Fig. 5D and E). The antioxidant *sod-2* and the mitochondrial chaperones *hsp-6* and *hsp-*

*60* were also induced by deficiencies of *nuo-6* or *cco-1* [28, 42, 44]. Therefore, protective responses were motivated due to deficiencies of genes encoding essential mitochondrial proteins, which was reminiscent of the scenario of PQ treatments. It is conceivable that some of the feedback responses such as the up-regulation of antioxidant enzymes may also contribute to ROS regulation in the context of mitochondrial dysfunctions. The data shown here were in contrast to the report that ROS, especially mitochondrial ROS, were up-regulated by deficiencies of *nuo-6* or *isp-1* [13]. However, Yee et al did not show fluorescent images of worms stained with ROS probing dyes. Interesting, the same research group in an earlier study reported decreased ROS levels in the same mutant strains. It should be noted that the lysis procedure would generate extra ROS and the variations of data obtained from isolated mitochondria were larger and the results were less reproducible, comparing to that obtained by directly staining in living worms (our observation).

## 4. Discussion

4.1. ROS levels are determined by both pro- and anti-oxidant capacities

Our results suggested that chronic prooxidant stresses would persistently and excessively activate the antioxidant responses and led to lower ROS and enhanced adaption to adverse conditions in the long term. Therefore, ROS should be determined by both the pro- and anti-oxidant capacities as illustrated in Figure 6. In support of this view, increased ROS levels due to increased mitochondrial metabolism and respiration activated antioxidant enzymes and led to further decrease of ROS [35].

However, recent studies reported increased ROS, especially mitochondrial ROS, in worms that were treated chronically with PQ or were deficient in *isp-1* or *nuo-6* [13]. The differences in approaches of measuring ROS should explain the discrepancy. In previous studies fluorescent images of worms stained with ROS or superoxide probing dyes were not shown, and ROS were measured in isolated mitochondria using flow cytometry technique [13]. However, there are many problems associated with in vitro measurement [15, 21]. For example: (i) Isolated mitochondria are exposed to oxygen levels that are 4–5 times higher; (ii) In the cells, mitochondria are tightly connected with cytoskeleton and other organelles such as the endoplasmic reticulum, which contribute to its normal function and ROS production; (iii) Substrates and ADP provided for isolated mitochondria are at saturating concentrations that would never be encountered in vivo; (iv) Worm lysis leads to intracellular or intraorganelle release of transition metal ions such as ferrous, which further catalyzes the generation of free radical through Fenton reaction. Therefore, ROS measured in isolated mitochondria were un-physiological and the results obtained using this approach cannot be directly extrapolated to the situation in vivo. It was thus recommended to measure ROS in living worms instead of in worm lysates [15, 21]. In addition, we found that worms heat-shocked at 37°C for only 30 minutes had dramatically increased ROS levels compared to control, suggesting the high sensitivity of ROS to perturbation of environmental factors. More procedures and time are required during in vitro measurement, thus variations of ROS data obtained are larger. At least the fluorescent images of worms stained with the probing dyes should be shown, considering that reliable ROS data are prerequisite for getting any scientific conclusions.

We were unable to found the elevation of ROS upon chronic PQ treatments or RNAi knockdown of *nuo-6*. It should be noted that mild reduction of mitochondrial ROS in worms with mutations in *nuo-6* or *isp-1* was reported [12]. ROS measurement in living worms should be reliable, because we indeed found elevated ROS in aged worms and in worms treated with low dose of ultraviolet irradiation or short term of heat shock. And we also observed elevated ROS after transient PQ treatments.

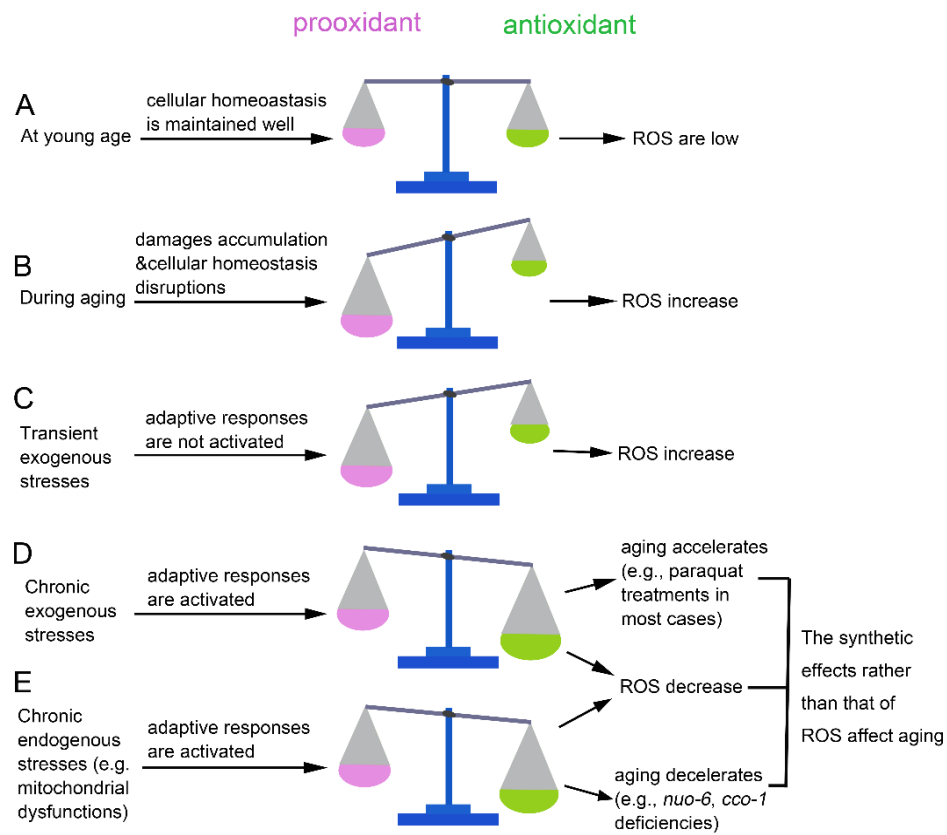

**Fig. 6. A model illustrating the interactions between prooxidant stresses, ROS, and aging.** (A) At young age cellular homeostasis is maintained well, the prooxidant and antioxidant capacities are well balanced, and ROS are low. (B) During aging the prooxidant capacity overshadows the antioxidant capacity and ROS gradually increase as the result of cellular damages accumulation and cellular homeostasis disruptions. (C) In response to transient exogenous stresses, the prooxidant capacity elevates immediately but the antioxidant capacity does not. In this scenario ROS increase in short term. (D, E) Under chronic exogenous or endogenous stresses adaptive responses are activated and ROS decrease in the long term. But aging accelerates under PQ treatments in most cases and decelerates by knocking down of *nuo-6* or *cco-1*, suggesting that the synthetic effects rather than that of ROS affect aging.

4.2. The pro-longevity effects, if any, should be ascribed to the synthetic effects of secondary responses rather than that of ROS

Based on these results and previous reports it can be concluded that ROS elevation is not associated with slowing down of aging (Fig. 6). Even if proper concentration of PQ still increased longevity when growth retardation was taken into consideration, the beneficial effect of which should not be ascribed to ROS. It is not only because ROS are decreased under chronic stresses, but also because multiple secondary responses that affects longevity are activated. If some of the secondary responses are pro-longevity then ROS should be the opposite because the formers are motivated by and counter against the latter's rising. Consistently, numerous studies showed that some of the antioxidants including resveratrol, N-acetylcysteine, vitamin C, reduced glutathione, and thioproline were able to increase longevity, and over-expression of the main superoxide dismutase *sod-1* also significantly extended lifespan in *C. elegans.* We therefore have reasons to question the current view that ROS are beneficial for longevity. Considering that growth retardation is quite common and is usually accompanied with sickness or fragility in worms [45, 46], we also propose a suggestion that sometimes the extension of the post growth span rather than the whole lifespan should be regarded as evidence for "real" slowing down of aging.

## Conflict of interests

The authors declare that there is no conflict of interest.

## Funding sources

This work was supported by the National Natural Science Foundation of China (No. 81200253 and 81570760), the National Key Research and Development Program of China (No. 2016YFA0102200, 2017YFA0103900, and 2017YFA0103902), One Thousand Youth Talents Program of China to C. Zhang, the Program for Professor of Special Appointment (Eastern Scholar) at Shanghai Institutions of Higher Learning (No. A11323), the Shanghai Rising-Star Program (No 15QA1403600), and the Fundamental Research Funds for the Central Universities of Tongji University.

## Author contributions
Y.R. conceived this study; Y.R., S.C., M.M., Congjie Z., K.W., F.L., W.G., and J.H. performed the experiments; Y.R., S.C., M.M., Congjie Z., and K.W. contributed to the analysis and interpretation of data; Y.R. wrote the manuscript; Y.R. and Chao Z. provided reagents and instruments for this study; Chao Z. revised and commented on the manuscript.

## Acknowledgement

We thank Dr Hezhi Fang in Wenzhou Medical University for technical assistance, and Drs. Jianxin Lu, Wei Li, and Zhifa Shen in Wenzhou Medical University for comments on the manuscript. Some strains were provided by the Caenorhabditis Genetics Center (CGC), which is funded by NIH Office of Research Infrastructure Programs (P40 OD010440).## References

[1]     D. Harman, Aging: a theory based on free radical and radiation chemistry, J Gerontol. 113 (1956) 298-300.

[2]     C.F. Labuschagne, E.C. Stigter, M.M. Hendriks, R. Berger, J. Rokach, H.C. Korswagen, A.B. Brenkman, Quantification of in vivo oxidative damage in Caenorhabditis elegans during aging by endogenous F3-isoprostane measurement, Aging cell. 122 (2013) 214-223.

[3]     S. Ayyadevara, M.R. Engle, S.P. Singh, A. Dandapat, C.F. Lichti, H. Benes, R.J. Shmookler Reis et al., Lifespan and stress resistance of Caenorhabditis elegans are increased by expression of glutathione transferases capable of metabolizing the lipid peroxidation product 4-hydroxynonenal, Aging cell. 45 (2005) 257-271.

[4]     R.S. Balaban, S. Nemoto, T. Finkel, Mitochondria, oxidants, and aging, Cell. 1204 (2005) 483-495.

[5]     B.P. Hubbard, D.A. Sinclair, Small molecule SIRT1 activators for the treatment of aging and age-related diseases, Trends in pharmacological sciences. 353 (2014) 146-154.

[6]     K.S. Bhullar, B.P. Hubbard, Lifespan and healthspan extension by resveratrol, Biochimica et biophysica acta. 18526 (2015) 1209-1218.

[7]     N. Moroz, J.J. Carmona, E. Anderson, A.C. Hart, D.A. Sinclair, T.K. Blackwell, Dietary restriction involves NAD(+) -dependent mechanisms and a shift toward oxidative metabolism, Aging cell. 136 (2014) 1075-1085.

[8]     D.E. Shore, C.E. Carr, G. Ruvkun, Induction of cytoprotective pathways is central to the extension of lifespan conferred by multiple longevity pathways, PLoS genetics. 87 (2012) e1002792.

[9]     J. Kim, N. Ishihara, T.R. Lee, A DAF-16/FoxO3a-dependent longevity signal is initiated by antioxidants, BioFactors. 402 (2014) 247-257.

[10]    A. Miranda-Vizuete, J.C. Fierro Gonzalez, G. Gahmon, J. Burghoorn, P. Navas, P. Swoboda, Lifespan decrease in a Caenorhabditis elegans mutant lacking TRX-1, a thioredoxin expressed in ASJ sensory neurons, FEBS letters. 5802 (2006) 484-490.

[11]    Y. Wang, S. Hekimi, Mitochondrial dysfunction and longevity in animals: Untangling the knot, Science. 3506265 (2015) 1204-1207.

[12]    W. Yang, S. Hekimi, A mitochondrial superoxide signal triggers increased longevity in Caenorhabditis elegans, PLoS biology. 812 (2010) e1000556.

[13]    C. Yee, W. Yang, S. Hekimi, The intrinsic apoptosis pathway mediates the pro-longevity response to mitochondrial ROS in C. elegans, Cell. 1574 (2014) 897-909.

[14]    J.M. Van Raamsdonk, S. Hekimi, Superoxide dismutase is dispensable for normal animal lifespan, Proceedings of the National Academy of Sciences of the United States of America. 10915 (2012) 5785-5790.

[15]    A. Sanz, Mitochondrial reactive oxygen species: Do they extend or shorten animal lifespan?, Biochimica et biophysica acta. 18578 (2016) 1116-1126.
16

## Supplemental materials

**Fig. S1. Egg hatching started juglone treatments reduced ROS levels.** ROS were measured in 6.5-day old N2 worms treated with corresponding concentrations of juglone from egg hatching (t=0 at egg hatching). Data are presented as mean ± SEM, and n represents number of worms.

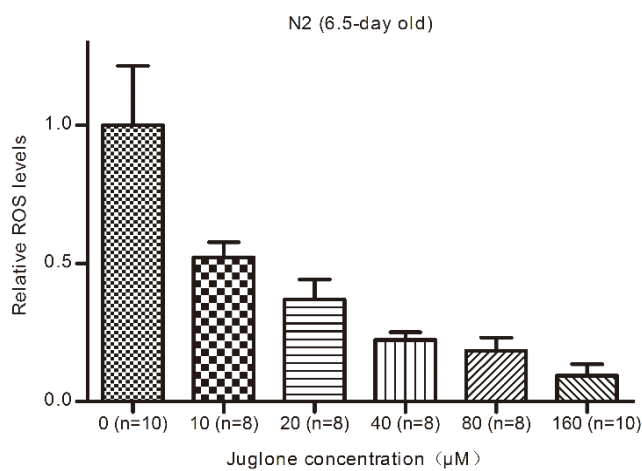

**Table S1. Detailed summary of lifespan data**

| PQ (mM) | Starting point of treatments | Mean lifespan ±SEM | N | Significance level | % of control lifespan |
|---|---|---|---|---|---|
| Figure. 2A | | | | | |
| 0 | Egg hatching | 24.79±0.52 | 53 | - | 100 |
| 0.03 | Egg hatching | 25.44±1.00 | 35 | n.s. | 103 |
| 0.05 | Egg hatching | 23.22±0.63 | 54 | n.s. | 91 |
| 0.1 | Egg hatching | 23.52±0.58 | 62 | n.s. | 95 |
| 0.2 | Egg hatching | 24.58±0.57 | 50 | n.s. | 99 |
| 0.3 | Egg hatching | 26.62±0.71 | 44 | ** | 107 |
| 0.5 | Egg hatching | 24.67±0.88 | 36 | n.s. | 100 |
| Figure. 2E | | | | | |
| 0 | Young adult | 22.80±0.62 | 40 | - | 100 |
| 0.03 | Young adult | 18.88±0.83 | 45 | ** | 83 |
| 0.05 | Young adult | 16.52±0.85 | 44 | *** | 72 |
| 0.1 | Young adult | 18.49±0.85 | 65 | ** | 81 |
| 0.2 | Young adult | 16.89±0.74 | 55 | *** | 74 |
| 0.3 | Young adult | 16.80±0.60 | 51 | *** | 74 |
| 0.5 | Young adult | 16.29±0.25 | 52 | *** | 71 |

All lifespan tests were carried out at 20°C. OP50 was used as the food source. All survival analyses were done with Graphphad Prism 5.0 software using Kaplan Meier analysis and log-rank test. N represents total worms, and $P \leq 0.05$ is considered statistically significant.

**Table S2. Primer sequences for quantitative RT-PCR.**

| Primer name | Sequence |
|---|---|
| *act-1* Forward | 5'-CCGCTCTTGCCCCATCAAC-3' |
| *act-1* Reverse | 5'-AAGCACTTGCGGTGAACGATG-3' |
| *hsp-6* Forward | 5'-CAGAGGTTCAAAAGGACTTAAAGG-3' |
| *hsp-6* Reverse | 5'-TGGAACTGTAACAACGGCG-3' |
| *sod-3* Forward | 5'-GATGGACACTATTAAGCGCGA-3' |
| *sod-3* Reverse | 5'-GTTTGCACAGGTGGCGATC-3' |
| *gst-4* Forward | 5'-CAAAGCTGAAGCCAACGACT-3' |
| *gst-4* Reverse | 5'-GAATTGACGGAAAAAGAATATGAA-3' |
| *hsp-16.2* Forward | 5'-CTCCATCTGAGTCTTCTGAGATTGT-3' |
| *hsp-16.2* Reverse | 5'-GGTAGAAGAATAACACGAGAAAATG-3' |